\documentstyle[aps,epsfig,lineno,amsmath,hhline,multirow,array]{revtex}

\newcommand{\ffig}[4]{\begin{figure}[hp]\vfill\begin{center}
\mbox{\epsfig{figure=#1,height=#2}}\caption{#3}\label{#4}
\end{center}\vfill\end{figure}}

\newcommand{\PO}{\rm l \! P }
\newcommand{\RO}{\rm l \! R }
\newcommand{\xpom}{x_{\PO} }

\begin{document}

\title{Pomeron structure functions from HERA to Tevatron and LHC}
\author{C. Royon\thanks{%
 Service de Physique des Particules,
CE-Saclay, F-91191 Gif-sur-Yvette Cedex, France},
L. Schoeffel \thanks{%
 Service de Physique des Particules,
CE-Saclay, F-91191 Gif-sur-Yvette Cedex, France},
R.Peschanski \thanks{%
 Service de Physique Th\'eorique, Unit\'e de recherche 
associ\'ee 
au CNRS, 
CE-Saclay, F-91191 Gif-sur-Yvette Cedex, France},
E.Sauvan \thanks{%
 CPPM, F-13288 Marseille Cedex 09, France
Marseille, France}
}
\maketitle

\begin{abstract}
The proton diffractive structure function $F_2^{D(3)}$ measured in the H1 and 
ZEUS experiments at HERA are analysed in terms of perturbative QCD in the 
perspective of the QCD extrapolation to the Tevatron and the LHC. It is 
shown that both data sets can be well described by a QCD analysis
in which point-like parton distributions, evolving according to the 
next-leading DGLAP equations, are assigned to the leading and sub-leading Regge 
exchanges. For present data from H1 and ZEUS the gluon distributions are found to 
be quite different and we give the corresponding sets of quark and gluon 
parton distributions for the Pomeron, extracted from the two experiments. An 
extrapolation to the Tevatron range is compared with CDF data on single 
diffraction. Conclusions on factorization breaking between HERA and Tevatron  
critically depend on whether H1 (strong violation) or ZEUS (compatibility 
at low $\beta$) fits are taken into account. Using the  double Pomeron 
formulation  in central diffractive dijet production we show that the Tevatron 
mass fraction is much sensitive to the high $\beta$ tail of the gluon in the 
Pomeron, suggesting a new way of handling the otherwise badly known  gluon 
distribution in the Pomeron. Extrapolation of the fits to very high $Q^2$ are 
given since they will be relevant for QCD and diffraction studies at the LHC.
\end{abstract}

\section{Introduction}

Since years, the Pomeron remains a subject of many interrogations. Indeed,
defined as the virtual colourless carrier of strong interactions, the nature 
of the Pomeron is still a real challenge. While in the perturbative
regime  of QCD it can be defined  as a compound system of two perturbatively 
Reggeized gluons\cite{bfkl} in the approximation of resumming the leading logs in 
energy, its non-perturbative structure is basically unknown.

However, in the recent years, an interesting experimental investigation on
``hard'' diffractive processes led to a new insight into Pomeron problems. 
It is now experimentally well established at HERA \cite{f2d94,zeus,zeusnew} 
that a substantial fraction of $ep$ events is contributable to diffraction,
i.e. color singlet exchange, initiated by a highly virtual photon. Starting with 
the pioneering theoretical work of Ref.\cite{ingelman}, the idea of a point-like 
structure of the Pomeron exchange opens the way to the determination of its 
parton (quark and gluon) distributions, where the Pomeron point-like structure 
can be treated in a similar way as (and compared to) the proton one. Indeed, 
leading twist contributions to the proton diffractive structure functions can 
be defined by factorization properties \cite{soper} in much the same way as for 
the full proton structure functions themselves. As such, they should obey DGLAP 
evolution equations \cite{dglap}, and thus allow for perturbative predictions 
of their  $Q^2$ evolution.

In Ref.\cite{pap2001}, we have shown that the data are well described by a QCD 
analysisin which point-like parton distributions, evolving according to the DGLAP
equations, are assigned to the leading and sub-leading Regge exchanges. The gluon 
distributions were found to be quite different for H1 and ZEUS. Then, in 
Ref.\cite{pap2001}, we have derived sets of quark and gluon parton 
distributions for the Pomeron, and  predictions for the charm and the 
longitudinal diffractive structure function from the QCD fit. An extrapolation to 
the Tevatron range was also compared with CDF data on single diffraction. In 
Ref.\cite{pap2002}, we have asked the question whether the quark and gluon 
distributions in the Pomeron obtained from QCD fits to hard diffraction processes 
at HERA could be dynamically generated by QCD evolution from a state made of {\it 
valence-like}  gluons and sea quarks as an input. By a method combining {\it 
backward} $Q^2$-evolution for data exploration and {\it forward} $Q^2$-evolution 
for a best fit determination, we have found that the diffractive structure 
functions published by the H1 collaboration at HERA  could be  described by a  
simple {\it valence-like} input at an initial low scale. The same property was 
not achieved when using ZEUS data.

These previous analyses were based on diffractive deep inelastic scattering (DIS) 
cross-sections determined at HERA \cite{f2d94,zeus}. It can be seen as a first 
stage in studying aspects of the QCD fit technique applied to diffraction, as for 
example the flexibility of the parton distributions parameterisations. 

In the present paper, we produce new sets of diffractive parton distribution 
functions (DPDFs) extracted from the presently published data sets, which consist 
in the already mentionned set for H1 \cite{f2d94} and the latest results from 
ZEUS diffractive DIS cross-sections \cite{zeusnew}. The results on DPDFs are 
close to those of Ref.\cite{pap2001}. We give for the first time
a new method to evaluate the uncertainty on the gluon density at
large values of $\beta$, which is a key point for any further discussion
of the influence of DPDFs at Tevatron or LHC.

Then, we present the main result of this article, which is to
show how these diffractive parton densities 
enter in the simulations of the dijet
mass fraction at the Tevatron in 
central diffractive dijet production within the double Pomeron formulation.
We show that this quantity is
much sensitive to the high $\beta$ tail of the gluon in the Pomeron. 
Our analysis
allows to provide the extrapolation of the fits to very high $Q^2$ since they
will be relevant for QCD and hard diffraction  studies at the LHC.

\section{Extraction of parton distributions in the Pomeron and QCD fits}

\subsection{Formulation}

It is well known \cite{f2d94,pap2001} that the diffractive structure function 
$F_2^{D(3)}$, measured from DIS events with large rapidity gaps, can be 
investigated in the framework of Regge factorization and expressed as a sum of 
two factorized contributions corresponding to a Pomeron and secondary Reggeon 
trajectories :

\begin{eqnarray}
F_2^{D(3)}(Q^2,\beta,x_{\PO})=
f_{\PO / p} (x_{\PO})\ F_2^{\PO} (Q^2,\beta)\ 
+ \ f_{\RO / p} (x_{\PO})\ F_2^{\RO} (Q^2,\beta) \ .
\label{reggeform}
\end{eqnarray}

In this parameterisation, $F_2^{\PO}$ can be interpreted as the Pomeron structure 
function  and $F_2^{\RO}$ as an effective Reggeon structure function, with the 
restriction that it takes together into account various secondary Regge 
contributions which can hardly be separated. The Pomeron, $f_{\PO / p}$, and 
Reggeon, $f_{\RO / p}$,  fluxes are assumed to follow a Regge behaviour with  
linear trajectories $\alpha_{\PO,\RO}(t)=\alpha_{\PO,\RO}(0)+\alpha^{'}_{\PO,\RO} 
t$, such that

\begin{equation}
f_{{\PO} / p,{\RO} / p} (x_{\PO})= \int^{t_{min}}_{t_{cut}}  {\rm d} t\ 
\frac{e^{B_{{\PO},{\RO}}t}}
{x_{\PO}^{2 \alpha_{{\PO},{\RO}}(t) -1}} 
\label{flux}
\end{equation}
where $|t_{min}|$ is the minimal kinematically-allowed value of $|t|$ and
$t_{cut}=-1$ GeV$^2$ is the limit of the measurement. The values of the $t$-slope 
parameters are taken from hadron-hadron data ($\alpha^{'}_{\PO}=0.26$ GeV$^{-2}$, 
$\alpha^{'}_{\RO}=0.90$ GeV$^{-2}$, $B_{\PO}=4.6$ GeV$^{-2}$, $B_{\RO}=2.0$ 
GeV$^{-2}$), since the data are not precise enough to determine them 
\cite{f2d94}.

We assign parton distribution functions to the Pomeron and to the Reggeon, 
considered as coumpound states of fundamental quarks and gluons. A simple and 
well-known prescription of QCD factorisation is that the parton distributions of 
both the Pomeron and the Reggeon are parameterized in terms of non-perturbative 
input distributions at some low scale which was adopted here to be $Q_0^2= 3$ 
GeV$^2$. The pion structure function \cite{GRVpion} is assumed to be also valid 
for the sub-leading Reggeon trajectory with a free global normalization to be 
determined by the data \footnote{We checked that changing
the pion structure function by some amount (20\%) does not change significantly
the parton distributions. For even better checking, another pion structure 
function \cite{owens} has also been used without any sizable modification of the 
results.}. 

For the Pomeron, a quark flavour singlet distribution,
$z{ {S}}(z,Q^2)=u+\bar{u}+d+\bar{d}+s+\bar{s}$, and a gluon distribution, $z{\it 
{G}}(z,Q^2)$, are parameterized in termsof coefficients $C_j^{(S)}$ and 
$C_j^{(G)}$ at $Q^2_0=3$ GeV$^2$ as it was done in Ref.\cite{f2d94} such that

\begin{eqnarray}
z{\it {S}}(z,Q^2=Q_0^2) &=& \left(
\sum_{j=1}^n C_j^{(S)} \cdot P_j(2z-1) \right)^2
\cdot \exp{\left(\frac{a}{z-1}\right)} \\
z{\it {G}}(z,Q^2=Q_0^2) &=& \left(
\sum_{j=1}^n C_j^{(G)} \cdot P_j(2z-1) \right)^2
\cdot \exp{\left(\frac{a}{z-1}\right)}
\label{gluon}
\end{eqnarray}
where $z=x_{i/I\!\!P}$ is the fractional momentum of the Pomeron carried by
the struck parton, $P_j(\zeta)$ is the $j^{th}$ member in a set of Chebyshev 
polynomials, chosen such that $P_1=1$, $P_2=\zeta,$ and $P_{j+1}(\zeta)=2\zeta 
P_{j}(\zeta)-P_{j-1}(\zeta)$. In the following, we present the results in terms 
of the overall quark density with the hypothesis that 
$u=\bar{u}=d=\bar{d}=s=\bar{s}$, which means that each light quark density is 
equal to $zS/6$.

A sum of $n=3$ orthonormal polynomials is used so that the input distributions are 
free to adopt a large range of forms for a given number of parameters. Any bias 
towards a particular solution due to the choice of the functional form of the 
input distribution is therefore minimized.  

In Ref.\cite{pap2001}, we have used the following fixed Pomeron intercepts for H1 
and ZEUS data sets, $\alpha_{\PO}(0) = 1.20 \pm 0.09 $ and $\alpha_{\PO}(0) = 1.13 
\pm 0.04$ respectively, as well as $\alpha_{\RO}(0)=0.62 \pm 0.03$ for the Reggeon 
intercept in case of H1 (there is no need for Reggeons for ZEUS data). In the 
following, we introduce the Pomeron intercept as a free parameter in the QCD fit. 

\subsection{Data sets}
Only data points with virtuality $Q^2 \ge 3$ GeV$^2$, diffractive mass at the 
photon vertex $M_X \ge 2$ GeV and rapidity $y \le 0.45$ are included in the fit in 
order to avoid large higher-twist effects and the region that may be most strongly 
affected by a non-zero value of the longitudinal over transverse ratio $R$. 

The fits include 179 data points for H1 data \cite{f2d94} with 8 parameters (3 for 
the sea quark density, 3 for the gluon density and 1 for the normalization of the 
Reggeon contribution and 1 for the Pomeron intercept). Thus, we have 171 degrees 
of freedom with a $\chi^2$ of $203.2$ using statistical errors only.
In Ref.\cite{pap2001}, only 161 data points were included in the QCD fit procedure 
since an additional cut for $\beta<0.65$ was considered in the analysis. 
In the new version of the fits, we gain in sensitivity
with a larger number of data points, in particular on the gluon density.
For example, the third parameter for the gluon density $C_3^{(G)}$ was
not constrained in Ref.\cite{pap2001} with a value of $0.01 \pm 0.04$ and
it is found to be zero in the new analysis. Then, for H1 data, 3 parameters
for quarks and 2 parameters for the gluon densties are enough to give a
proper description of the cross section measurements through the QCD fit
procedure. Also, new results and parameters obtained in Ref.\cite{pap2001}
are in perfect agreement for the quarks within the quoted 
statistical uncertainties.

For ZEUS data \cite{zeusnew}, the fits include 102 data points with 7 parameters 
(the parameter corresponding to the normalization of the Reggeon contribution is 
absent). We have 95 degrees of freedom with a $\chi^2$ of $118.4$ using 
statistical errors only. Note that ZEUS measurements are realised for diffractive 
mass at the proton vertex $M_Y < 2.3$~GeV and squared transfer-momentum $t < 
1$~GeV$^2$, whereas H1 gives diffractive cross sections for $M_Y < 1.6$~GeV and $t 
< 1$~GeV$^2$. We have converted ZEUS data points to the same $M_Y$ range by 
multiplying ZEUS values by the factor $0.77$ \cite{zeusnew}.
This factor takes into account the conversion from
$M_Y < 2.3$~GeV to $M_Y < 1.6$~GeV for the ZEUS measurements, which
reflects mainly the fact that the proton-dissociation is reduced
for the range of H1 data set. The uncertainty on this number is of course
large (of about $\pm 0.1$ \cite{zeusnew}), 
but we do not take it into account in the following as we do
not propose a common fit of H1 and ZEUS data sets, which would require
a relative normalisation error.

On Fig.\ref{f4}, structure function measurements from both data sets
are shown.
In this last figure, ZEUS data points have been 
multiplied by the factor 0.77 as explained above,
and extrapolated to H1 bin centers. This extrapolation is done just
for the plot, ZEUS data points have been included in the QCD fits at
their exact kinematic values.

To get solvable evolution equations, the parton distribution functions must 
approach zero as $z\rightarrow 1$. This is achieved by introducing in Eqs. 
\eqref{gluon} the exponential term with a positive value of the parameter $a$.  
Unless otherwise indicated, in the following fits $a$ is set to $0.01$ such that
this term only influences the parametrisation in the region $z>0.9$. This
term is only present to ensure the convergence and plays a r\^ole in a domain
where we do not include data (for instance, at $\beta=$0.65, its value
is equal to 0.97). 

The functions $z{\it{S}}$ and $z{\it{G}}$ are evolved to higher $Q^2$ using the
next-to-leading order DGLAP evolution equations (with the code of 
Ref.\cite{lolo}), with $\alpha_S(M_Z^2)=0.118$ (in Ref.\cite{pap2001}, a lower 
value of $\alpha_S(M_Z^2)$ has been considered).  The contribution to
$F_2^{I\!\!P}(\beta,Q^2)$ from charm quarks is calculated in the fixed flavour 
scheme using the photon-gluon fusion prescription given in Ref.\cite{ghrgrs} as 
implemented in Ref.\cite{rapgap}. The contribution from heavier quarks is 
neglected. 

No momentum sum rule has been imposed because of the theoretical uncertainty in
specifying the normalization of the Pomeron or Reggeon fluxes.It is also
not clear that such a sum rule is appropriate for the parton distributions of a 
virtually exchanged state. After fitting, however, we observed that the sum rules 
were fulfilled within 10\% despite being not required in the fits. 

\subsection{Results}

Already from Fig.\ref{f4} we can get a good idea of the relative gluon
densities between H1 and ZEUS data sets. Indeed, we notice a good
agreement at  $Q^2=7.5$~GeV$^2$ (and $\beta<0.65$) whereas
both sets of measurements show discrepancies at larger $Q^2=28.$~GeV$^2$.
ZEUS cross section values are below H1
for this $Q^2$ value, which means that scaling
violations in case of ZEUS are lower. Thus, from this basic
observation, we expect a lower gluon density in case of ZEUS data
compared to H1.


The resulting parton densities of the Pomeron
extrapolated from the QCD fit procedure
are presented in Fig.\ref{f12b} 
(extrapolated till $10000$~GeV$^2$) and
the parameter values are given in table I.

A few comments are in order : 
\begin{itemize}
\item
 We notice that the quark contribution is smaller compared to the gluon one, and 
much smaller for H1 prediction, confirming the standard results exposed for 
example in Ref.\cite{pap2001}. 
On  Fig.\ref{f12b}, the uncertainties of
the parton distributions are not shown but they can be derived from table I,
giving typically a 25\% error for the gluon density. 
\item
 In the range of the measurements, essentially below $Q^2=75$~GeV$^2$ for 
H1\cite{f2d94} and below $Q^2=55$~GeV$^2$ for ZEUS\cite{zeusnew}, the agreement
between H1 and ZEUS for quark densities is reasonnable. \\
\item
 Gluon densities for both data sets is quite different, especially at large 
$\beta$. This last point has been 
addressed above from the simple observation of scaling violations and
is more extensively discussed below. 
\end{itemize}

The result of the fit  is presented in Fig.\ref{f4} together with the experimental 
values. The fit  is plotted only in the $\beta - Q^2$ bins considered for the QCD  
analysis. In this last figure, 
as mentioned above,
ZEUS data points have been converted to the same 
$M_Y < 1.6$~GeV domain as H1 (multiplying ZEUS measurements by a multiplicative 
factor 0.77 as explained above) and extrapolated to H1 bin centers. However, the 
QCD-fit procedure and extraction of parton distributions have been done exactly on 
ZEUS published data points \cite{zeusnew}. The extrapolation to H1 bin centers is 
applied only for the presentation of Fig.\ref{f4} and we have checked that these 
conversion factors are small (of the order a few per cent). We see on this figure 
the good agreement of the QCD fit with the data points, which supports the
validity of the description of the Pomeron in terms of partons following 
perturbative QCD dynamics, even if the accuracy of the gluon determination is 
still quite poor. This provides a very nice perspective for diffractive analyses 
of more recent data to be released, where better accuracy will lead to a more 
precise QCD analysis.

\begin{center}
\begin{tabular}{|c|c|c|} 
\hline
 parameters & H1 &  ZEUS  \\ 
\hline\hline
 $\alpha_{\PO}$ & \ 1.19\  $\pm$ 0.02\  & \ 1.13\   $\pm$ 0.02\  \\
\hline
 $C_1^{(S)}$ & \ 0.21\      $\pm$ 0.05\  & \ 0.38\    $\pm$0.02\  \\
 $C_2^{(S)}$ & \ 0.04\  $\pm$ 0.02\  & \ -0.03\   $\pm$ 0.01\  \\
 $C_3^{(S)}$ & \ -0.14\      $\pm$ 0.02\  & \ -0.11\   $\pm$ 0.01\  \\ \hline
 $C_1^{(G)}$ & \ 0.59\       $\pm$ 0.40\  & \ 0.39\      $\pm$ 0.10\  \\
 $C_2^{(G)}$ & \ -0.01\  $\pm$ 0.01\  & \ -0.36\  $\pm$ 0.05\  \\
 $C_3^{(G)}$ & 0.00   & \ 0.05\  $\pm$ 0.02\  \\
\hline
 $N_{IR}$  & \ 14.11\  $\pm$ \ 1.13\  &  0.00 \\
\hline
\end{tabular}
\end{center}
\vskip .5cm
\begin{center}
{Table I- Pomeron quark and gluon densities parameters.
Note that $C_3^{(G)}$ is found to be negligeable for H1 data. Also, as mentionned
above, there is no Reggeon contribution for ZEUS data. For H1 data, the parameters 
are close to those of Ref.\cite{pap2001} within the quoted statistical 
uncertainties, even if we expect some differences as it is explained in the text : 
for example, more data points are considered and some small changes are done in the QCD fit 
procedure.
}
\end{center}

\subsection{Large $z$ behaviour}

In order to analyze in more detail the large $z$ behaviour of the gluon 
distribution $z{\it {G}}(z,Q^2=Q_0^2)$ and give a rough estimate of the 
systematic error related to our parametrizations, we consider the possibility 
to change the ansatz \eqref{gluon} by a multiplicative factor $(1-z)^{\nu}.$ We 
shall consider $\nu$ to vary in the interval $\pm 1$ in order to allow for the 
still large indetermination of the gluon distribution. If we include this
multiplicative factor $(1-z)^{\nu}$ in the QCD fit analysis with statistical
errors, we derive a value of $\nu = 0.0 \pm 0.6$. Then, the variation 
in the interval $\pm 1$ considered here takes also into account a systematic
effect of the order of the statistical uncertainty, which is correct.
As it is explained more extensively below, the understanding of the
large $z$ behaviour is of essential interest for any predictions at 
the Tevatron or LHC in central dijets production. In particular, a
proper determination of the uncertainty in this domain of momentum is
necessary and the method we propose in this article is a first estimate,
that can be propagated to other measurements. 

Following Ref.\cite{lopez}, one knows that starting with a given large $z$ 
input function of the form $(1-z)^{\nu_0},$ the effect of leading order DGLAP 
evolution is to change the end-point exponent  by  
\begin{equation}
(1-z)^{\nu_0}\ \to\ (1-z)^{\nu_0-\frac 
{16}{33-2n_f} \ \log{\alpha_s(Q^2)}}\ ,
\label{nu}
\end{equation} 
where $n_f$ is the number of active flavors, up to a normalization factor. 

Hence, let us consider a given solution of the DGLAP evolution equation, such as 
our structure function $z{\it {G}}(z,Q^2),$ and its  large $z$ tail at $Q_0^2$  
parametrized by $(1-z)^{\nu_0}.$ Then, changing the high $z$ behaviour of the 
input function \eqref{gluon} by $(1-z)^{\pm\nu}$ will change the exponent 
$\nu(Q^2)$ by the same shift, namely 
$$\nu(Q^2)\to {\nu_0-\frac 
{16}{33-2n_f} \ \log{\alpha_s(Q^2)}}\pm\nu\ \equiv\ \nu(Q^2) \pm \nu\ .$$
Note that the same property is valid for the singlet quark distribution 
\cite{lopez}.

To be fully consistent with the NLO analysis,we estimated the effect of 
next-leading DGLAP evolution on the modifications of the high $z$ behaviour due to 
next-leading effects \cite{lopez1}. We found that the  next-leading corrections to 
the overall  high $z$ exponent $1/\nu (Q^2)$ is of the order of its inverse $1/\nu 
(Q^2) \le 20$~\% when $\nu(Q^2)$ varies in the kinematical interval in 
consideration. This correction is  negligeable in the conditions of our fits.



\ffig{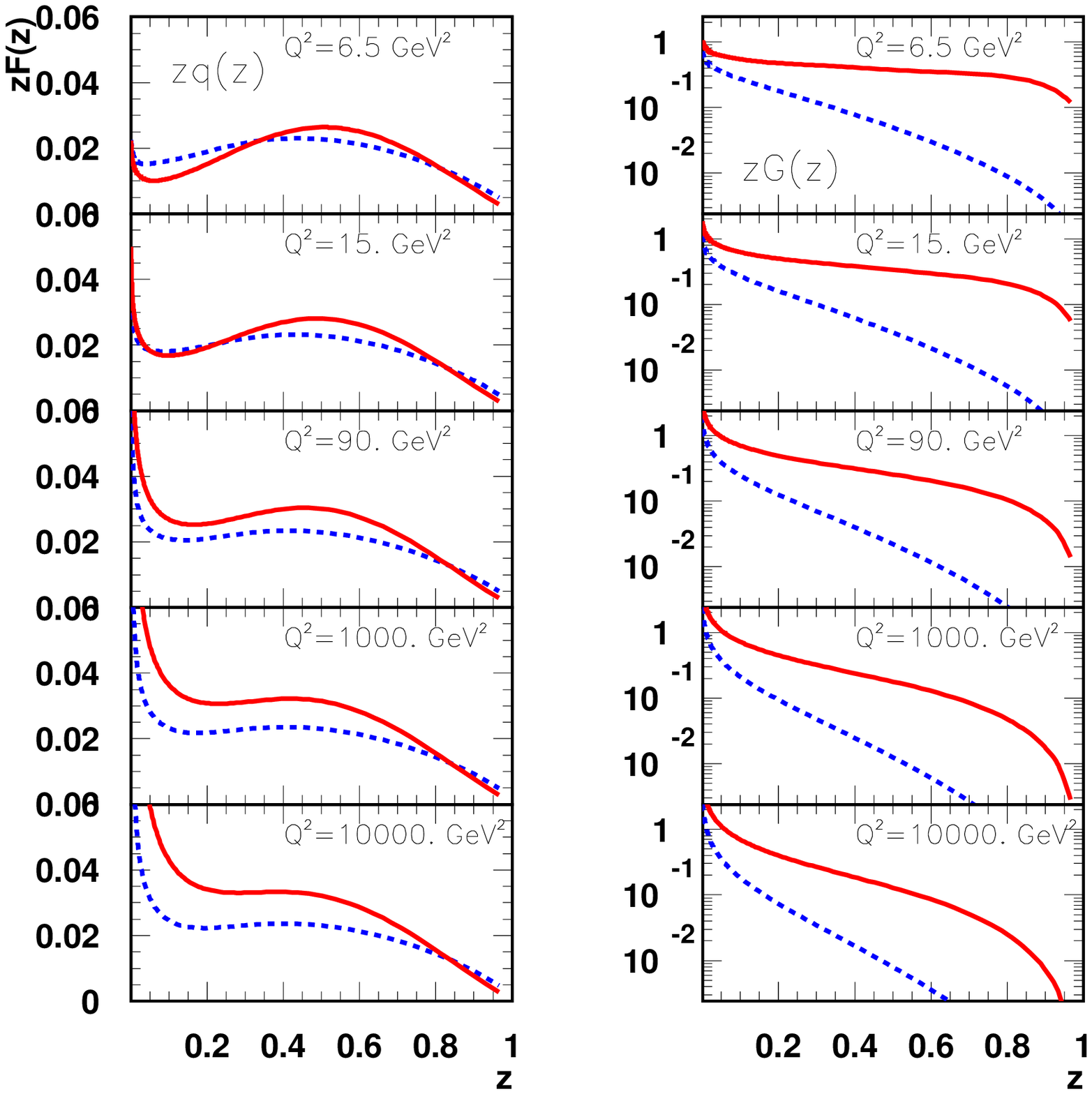}{160 mm}{
{\it Quark and Gluon distributions in the Pomeron.}(H1 : full line, ZEUS : dotted line)  
as a function of $z$, the fractional momentum of the Pomeron 
carried by the struck parton, from the fit on H1 and ZEUS data points with $Q^2 
\ge 3$ GeV$^2$. The parton densities are normalised to represent $\xpom$ times the 
true parton densities multiplied by the flux factor at $\xpom = 0.003$.
}
{f12b}

\ffig{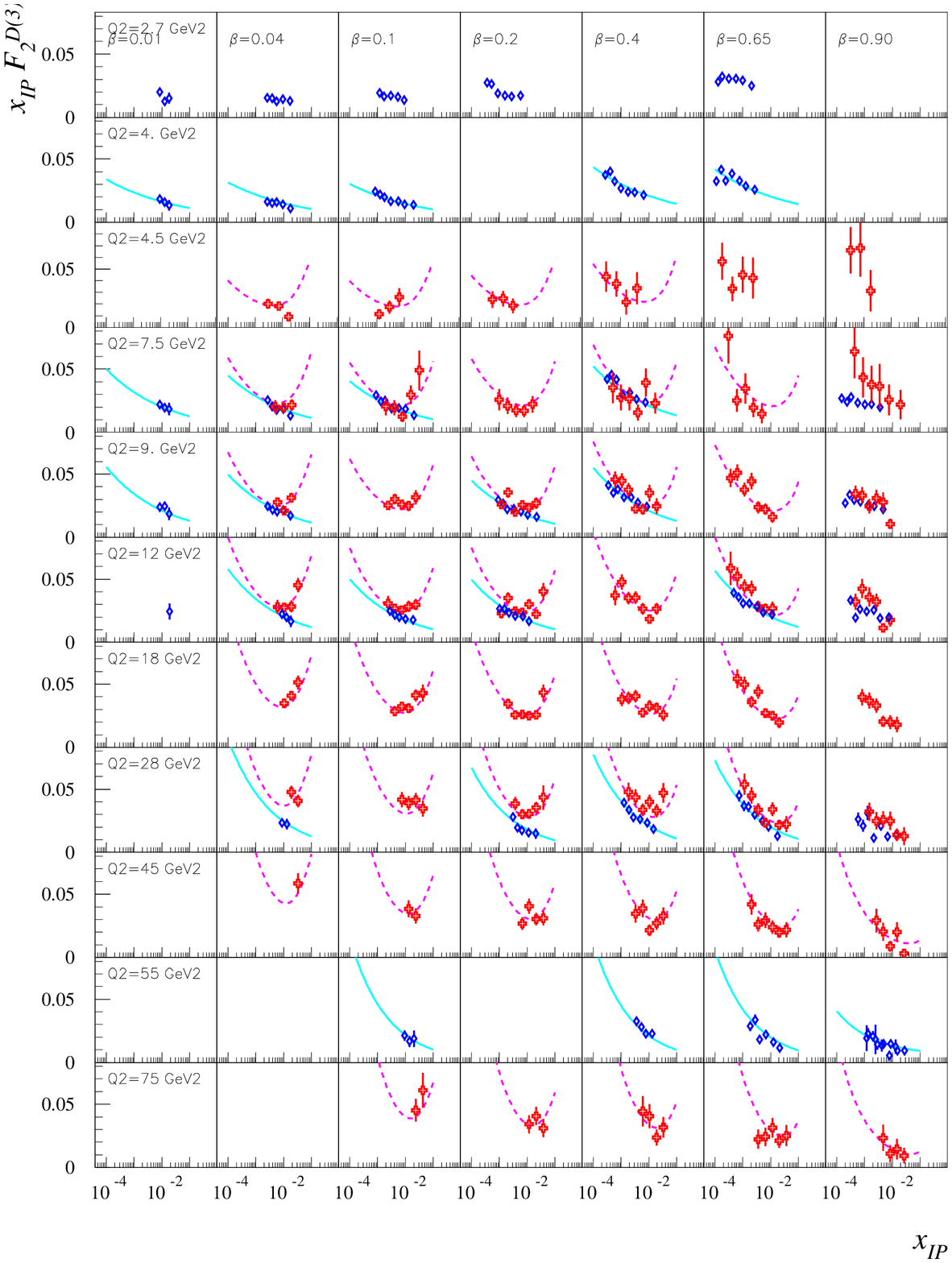}{200 mm}{
{\it Comparison of data with the results of the QCD fits.}
Data points: H1 (dashed line); ZEUS (full line).
QCD fits: H1 (crosses); ZEUS (diamonds). The ZEUS data points have been
converted to the same $M_Y$ range as H1 measurements and moved to the 
H1 bin centers (see text).
}
{f4}

\ffig{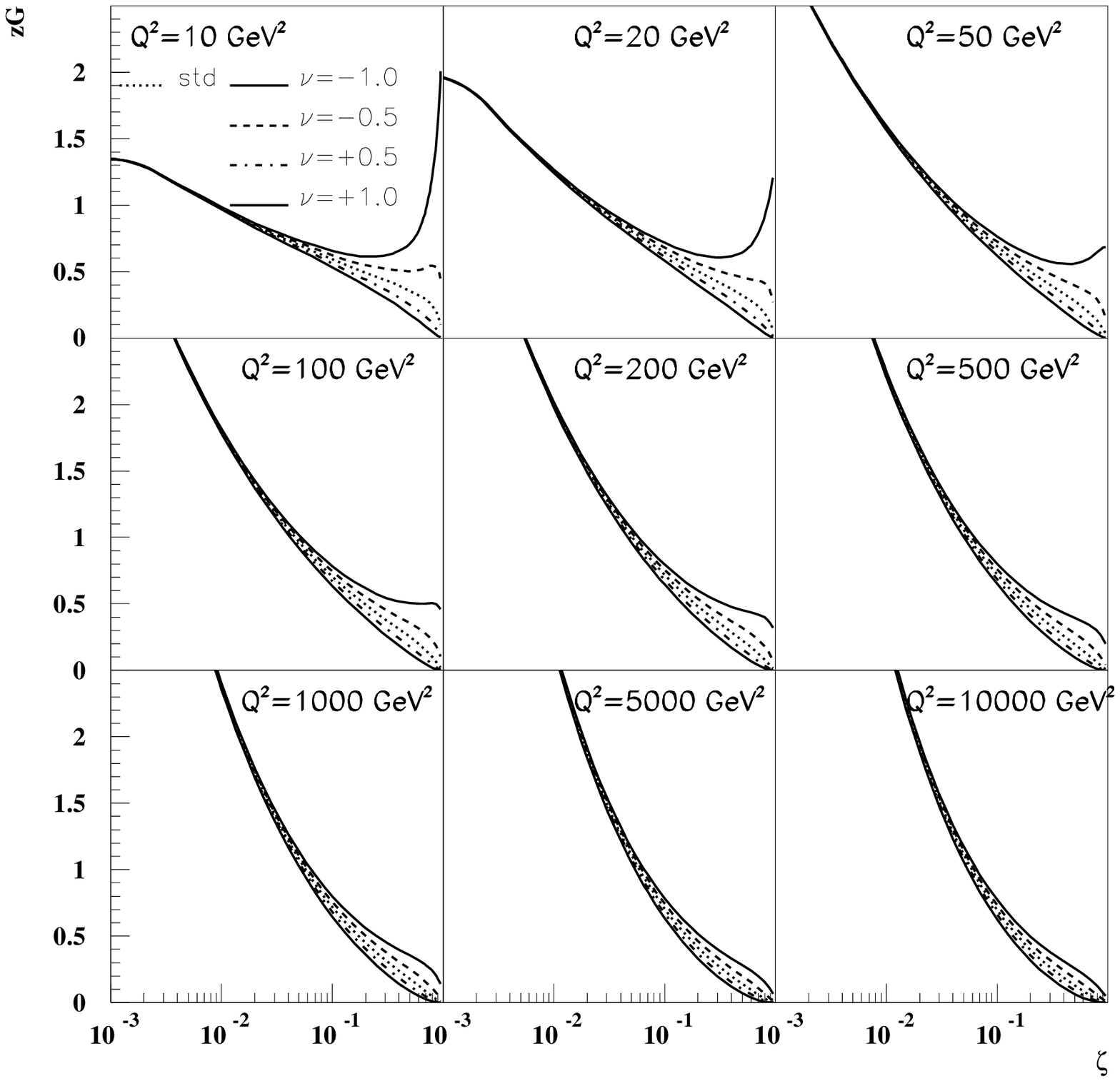}{90 mm}{
{\it Gluon distributions
of the Pomeron from the fit
on H1 data points alone with $Q^2 \ge 3$ GeV$^2$}. The parton densities are 
normalised to represent$\xpom$ times the true parton densities multiplied by the 
flux factor at$\xpom = 0.003$. A change by $\nu = \pm .5\ {\rm and}\ \pm 1$ have 
been performed (see text).
}
{fnu1}

\ffig{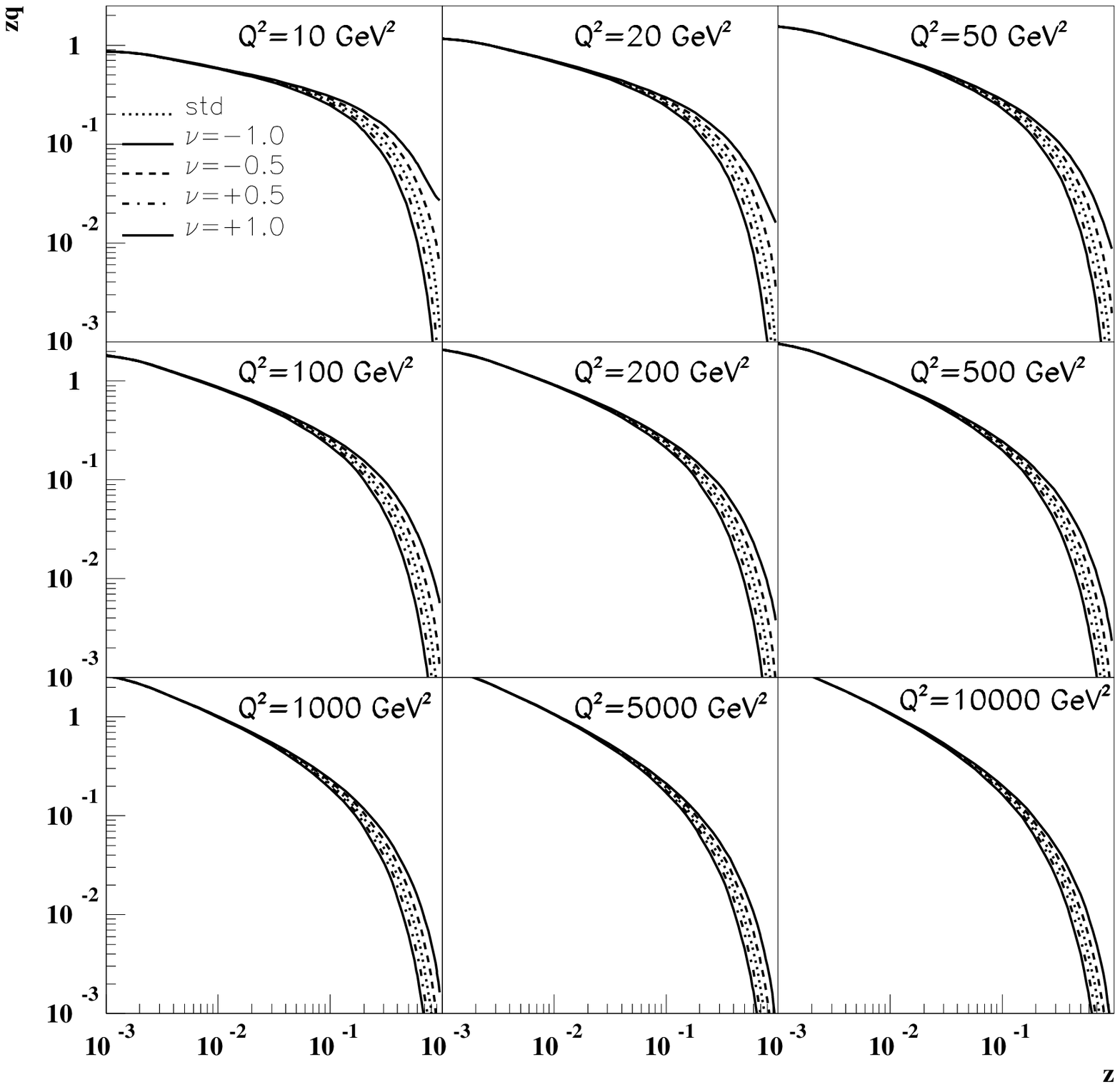}{90 mm}{
{\it Gluon distributions
of the Pomeron from the fit
on ZEUS data points alone with $Q^2 \ge 3$ GeV$^2$}. Same conventions as previous 
figure.
}
{fnu2}

\section{Extrapolation to Tevatron and comparison with CDF data}
The QCD fits we  obtained from HERA data allow us to make direct comparisons for 
measurements at the Tevatron. It is quite interesting to be able to  test directly 
the factorization breaking between HERA and the Tevatron using the measurements 
performed at both accelerators. We thus compare the extrapolations of the H1 and 
ZEUS QCD fits to the recent CDF single diffractive jet cross-section measurement 
\cite{cdf}. The result is given in Fig.\ref{cdf}. We note a large discrepancy both 
in shape and normalization between H1 predictions and CDF data, clearly showing 
factorization breaking. However, the ZEUS fits are more compatible in 
normalization with the CDF measurement even if the shape is not described 
properly. We know that the gluon density from ZEUS is between 2 and 3 times 
smaller than the one from H1. The predictions for single diffractive production of 
jets at the Tevatron are thus expected to be  different by an sizeable  factor 2 
to 3 between H1 and ZEUS since they correspond to single Pomeron exchange. If the 
large statistical and systematicuncertainties on the gluon density are taken into 
account (about 50\%for ZEUS, 25\% for H1), ZEUS data are compatible with  
factorization at low $\beta$. 

One should, however, question whether one has the right to extrapolate ZEUS 
results without introducing the additional Reggeon component. Namely the CDF 
measurement is in a region in $\xpom$ where the Reggeon contribution is important, 
in contrary to the ZEUS measurement. The error bar on the extrapolation from the 
fit for ZEUS is thus enhanced due to this uncertainty. Concerning the 
extrapolation of the H1 fit and the comparison with CDF data, one notices that the 
CDF data lie primilarly at low values of $\beta$ where there are few H1 data 
points. Moreover, in this kinematical domain, the H1 data have the tendency to lie 
more in the high $\xpom$ region where the Reggeon contribution is important and 
not well constrained by the fit. Thus, the extrapolation to the CDF domain suffers 
from large uncertainties. A combined fit using CDF data to constrain the low 
$\beta$ region and the HERA data to constrain the high $\beta$ domain would thus 
be of great interest.

More precise data from HERA and detailed comparisons with the Tevatron are thus 
needed to study precisely factorization breaking between both experiments. The 
discussion of eventual higher twist contributions is clearly also valuable in 
order to reach conclusions on the shape at larger $\beta.$ It is thus important to 
get an accurate measurement of the gluon density in the Pomeron at low values of 
$\beta$ from HERA. Furthermore the Forward Proton Detector \cite{FPD} installed by 
the D0 collaboration for Run II will be of great help to get a direct measurement 
of the diffractive structure functions.

\section{Possible measurements at the Tevatron}
The CDF and D\O\ experiments at the Tevatron can measure directly the dijet mass
fraction, defined as the ratio of the dijet mass to the total diffractive mass
measured either in roman pot detectors or in the main D\O\ or CDF detectors.

Let us first notice that we normalised the diffractive dijet cross section
using directly the CDF run I measurement previously which takes into account the
factorisation breaking between the Tevatron and HERA \cite{pap2001,cdf}.
It has already been mentionned that this measurement is quite sensitive to the
Pomeron structure in quark and gluons \cite{pap2001}. We display in 
Fig.\ref{fcdfc} the dijet mass fraction for the two different gluon distributions
described previously from H1 and ZEUS data after requiring two jets in the central 
detector with a transverse momentum greater than 25 GeV. We indeed see that the 
difference between the gluon distributions from H1 and ZEUS at HERA can be probed 
using Tevatron data, provided the survival gap probability (correction which is 
due to soft interactions) is constant over the full kinematical range. To better 
show the possibility of such a measurement, we give in Fig.\ref{fcdfb} the dijet 
mass fraction when the $\bar{p}$ is tagged in the dipole roman pot detector from 
the D\O\ collaboration. The dipole detectors show a good acceptance for $t$ close 
to $0.$ It would also be possible to require double tagged events (the antiproton 
in the dipole roman pot detector and the proton in the quadrupole one), but the 
acceptance of the quadrupole detectors only extends to $t \sim 0.8$ GeV$^2$ which 
cuts a lot of events.

On the other hand, it is also possible to probe the high-$\beta$ gluon density in
the Pomeron at the Tevatron again through the measurement of the dijet mass 
fraction (or the total diffractive mass). In Fig\ref{fcdfd}, we givethe dijet mass 
fraction using the shape of gluon distribution in the Pomeron coming
from H1. The sensitivity to the uncertainty on the gluon distribution at high
$\beta$ is indicated on that figure by multiplying the gluon distribution by $(1-
\beta)^{\nu}$, which enhances or decreases the high $\beta$ gluon distribution.
It is quite important to be able to constrain this distribution since it is a
direct background to an eventual exclusive signal at high $\beta$. 
We notice that it is indeed possible to constrain the high-$\beta$ gluon using 
Tevatron data. At the LHC, it will be possible to constrain better the 
high-$\beta$ gluon density using higher mass objects, for instance in $t \bar{t}$ 
inclusive diffractive production.

\ffig{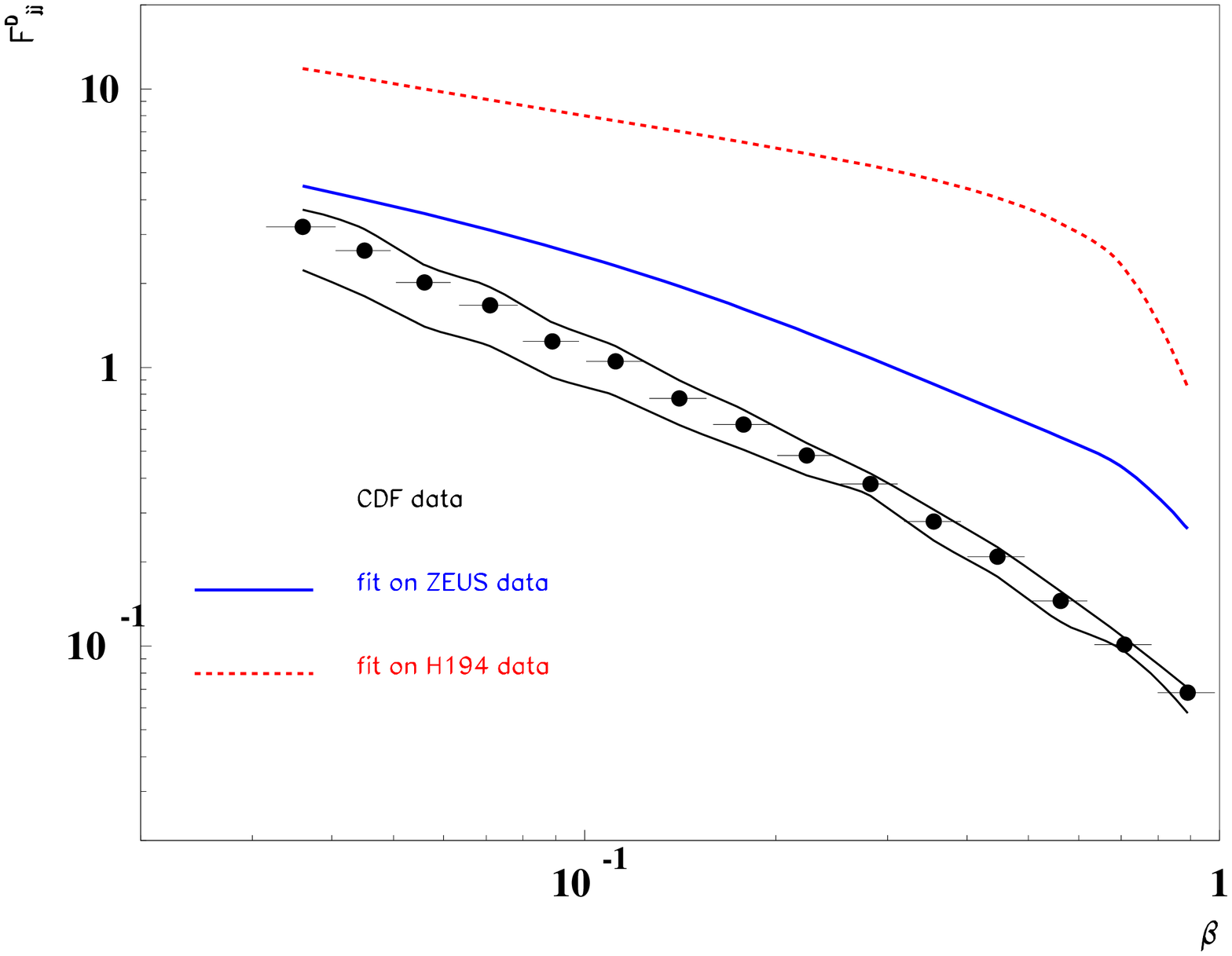}{95 mm}
{{\it $\beta$ distributions of CDF data  compared with extrapolations from the
diffractive parton densities. } The partons densities are extracted from  
$F_2^{D(3)}$ measurements by H1 and  by ZEUS collaborations}
{cdf}

\ffig{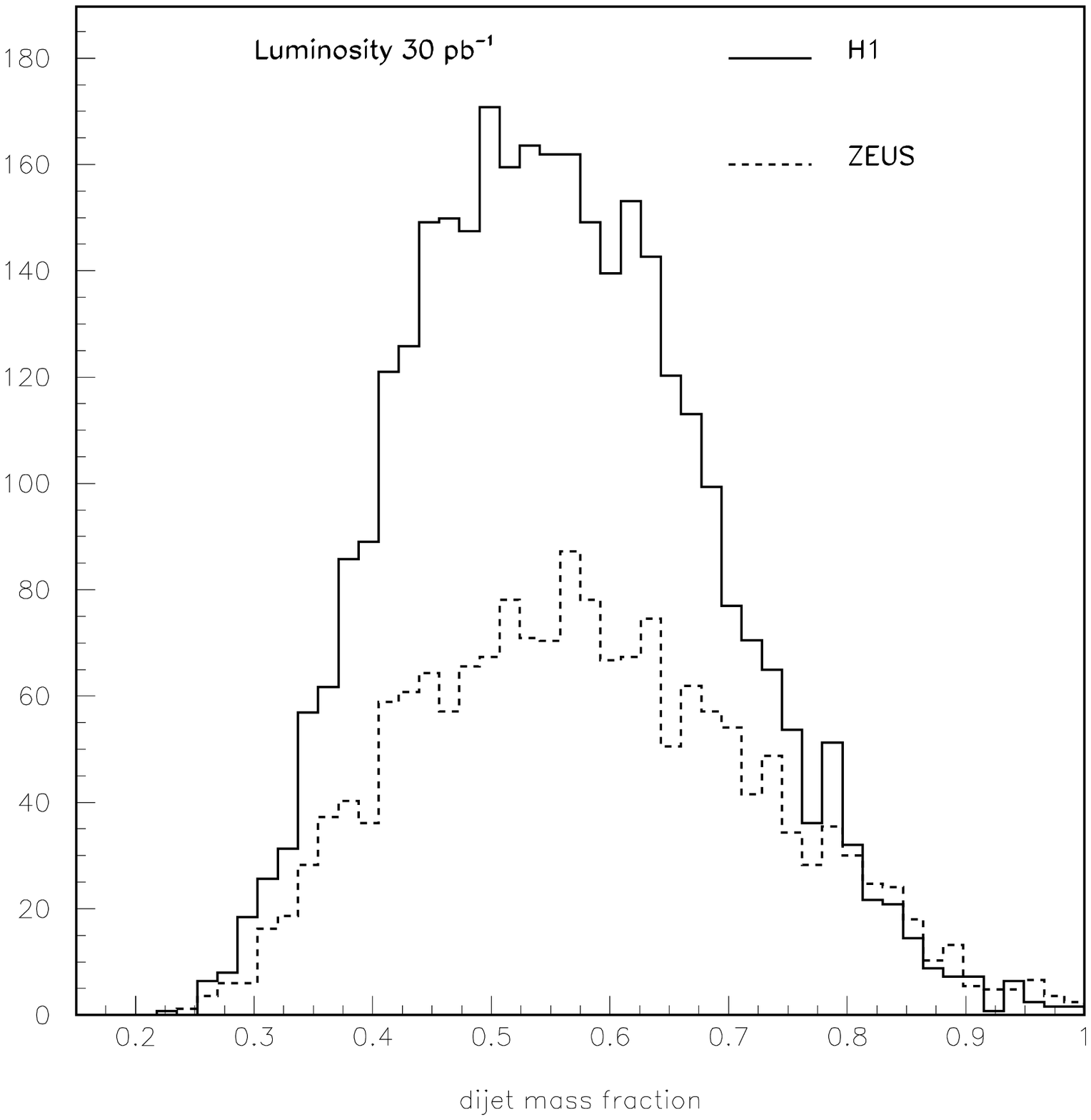}{95 mm}{
{\it Results for Dijet mass fraction at generator level.}  The H1 and ZEUS
parton distributions in the Pomeron serve as inputs (see text).}
{fcdfc}

\ffig{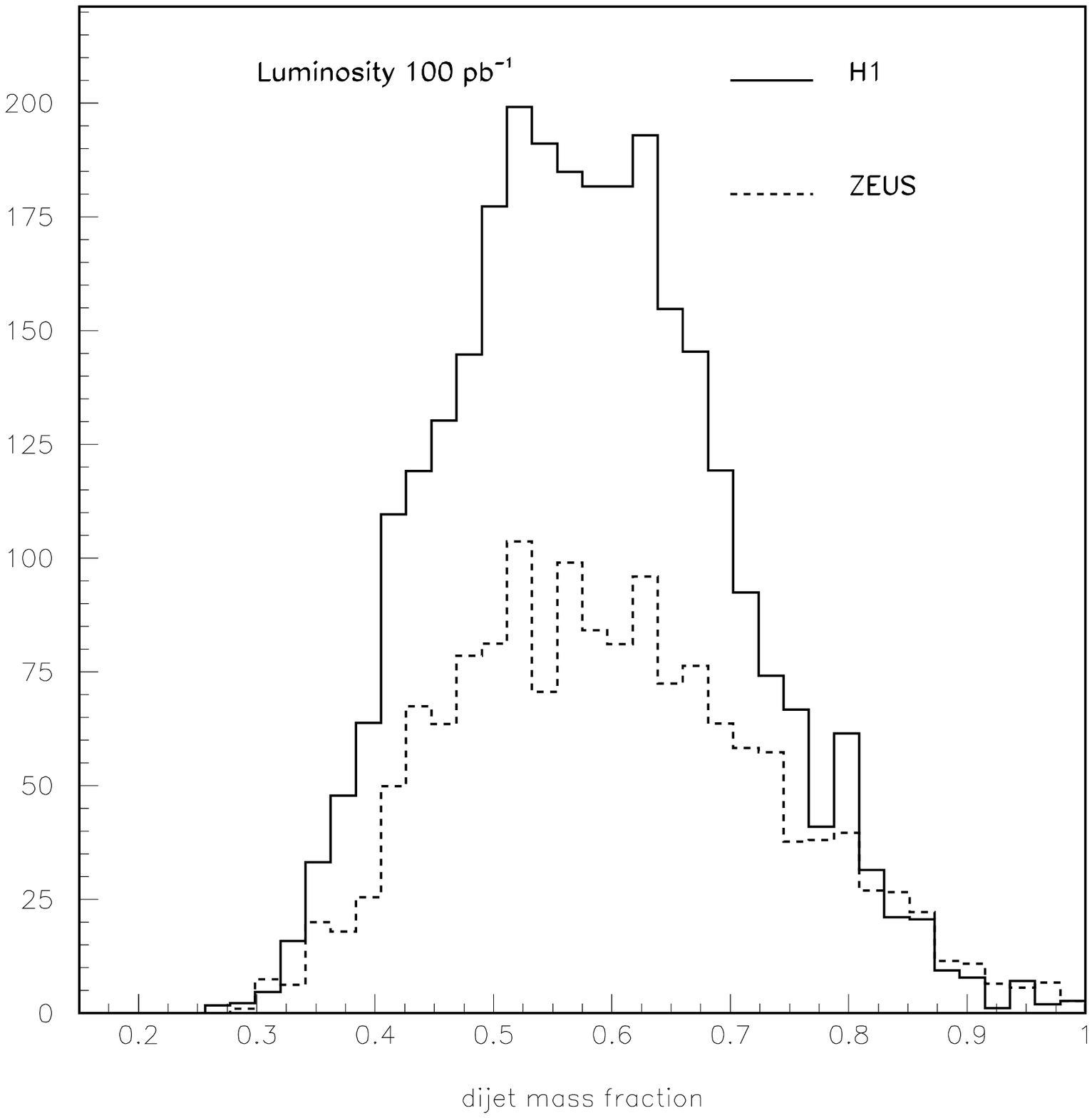}{95 mm}{
{\it Results for Dijet mass fraction in the D\O\ dipole roman pot acceptance at 
generator level.} One considers tagging a $\bar{p}$ (see text). Same convention as 
previous figure. Note that the CDF results would be similar.}
{fcdfb}

\ffig{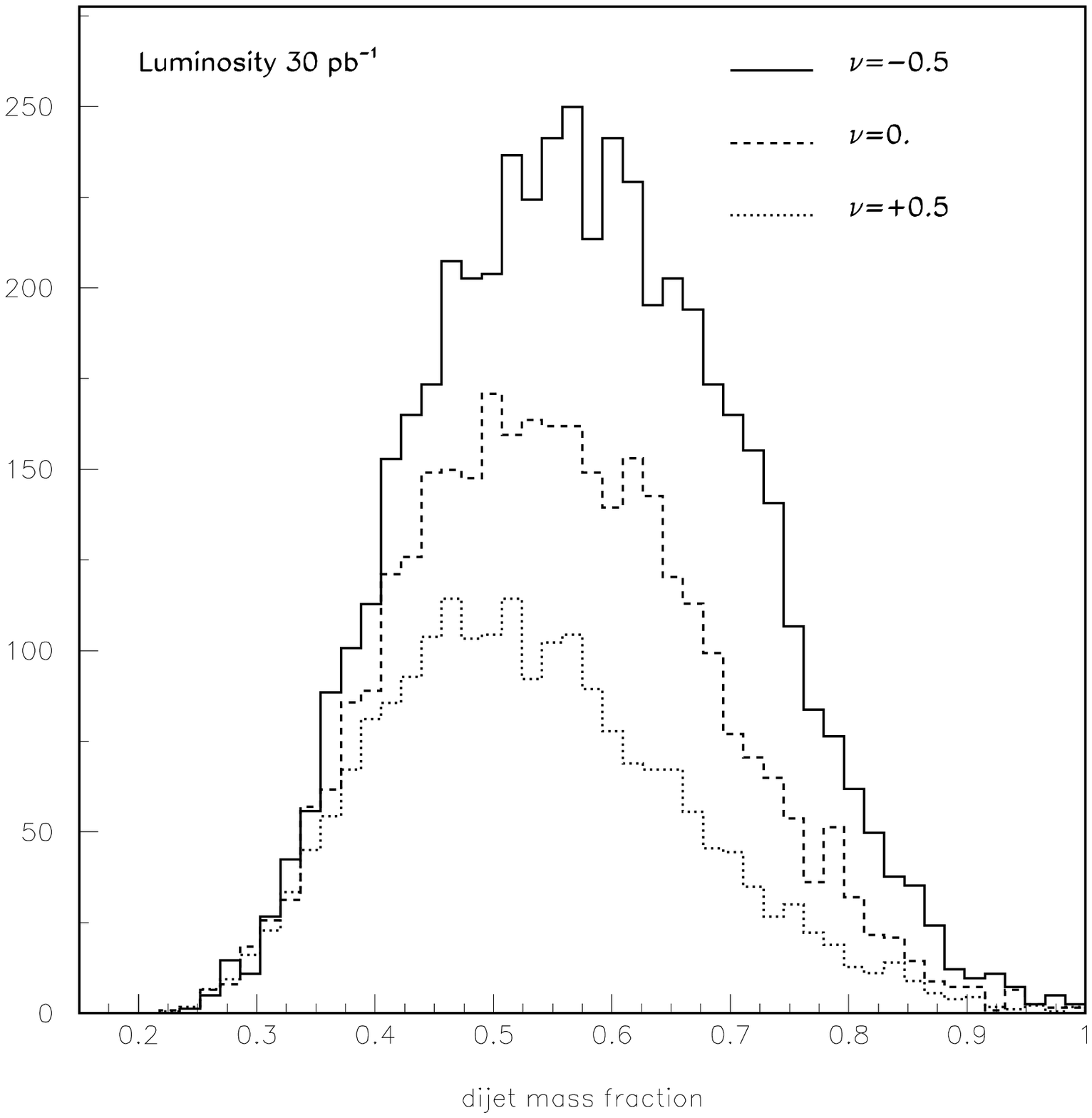}{95 mm}{
{\it Results for Dijet mass fraction for the H1 
parton distributions as a function of the high-$\beta$ tail.} The high-$\beta$ 
component of the gluon density in the Pomeron is modified ed by a factor $(1 - 
\beta)^{\pm \nu}$ to show the dependence of the dijet mass fraction on the 
parameter $\nu$. (see text).}
{fcdfd}

\section{Conclusion} 

We have shown that the proton diffractive structure function $F_2^{D(3)}$ 
measured in the H1 and ZEUS experiments at HERA can be well described by a 
perturbative  QCD analysisin which fundamental quark and gluon  distributions, 
evolving according to the next-leading DGLAP equations, are assigned to the 
leading and sub-leading Regge exchanges. 

The gluon distributions have been found to be quite different when extracted from 
present published data for H1 and ZEUS. An extrapolation to the Tevatron range 
have been compared with CDF data on single diffraction,leading at this stage to 
quite different conclusions on factorization breaking between HERA and 
Tevatron depending on whether H1 (strong violation) or ZEUS (compatibility 
at low $\beta$) fits are used. The discrepancies of the gluon densities between
both data sets have also been explained when looking simply at the scaling
violations, the QCD fit procedure giving a quantitative estimate of these
differences.
We have presented a new method to evaluate the uncertainty on the gluon density at
large values of $\beta$, which is a key point for any further discussion
of the influence of DPDFs at Tevatron or LHC.

In double diffractive dijet production in the central rapidity region, within the 
double Pomeron formulation, we have found that the Tevatron mass fraction  is very 
sensitive to the high $\beta$ tail of the gluon in the Pomeron. Extrapolation of 
the fits to very high $Q^2$ have been given since they will be relevant for QCD 
studies at the LHC.

Hence, we have shown that it is possible to constrain the high-$\beta$ gluon using 
Tevatron data, which may be of decisive interest for any future
measurement in central diffractive dijet topologies.

At the LHC, it will be possible to constrain better the high-$\beta$ 
gluon density using higher mass objects, for instance in $t \bar{t}$ inclusive
diffractive production. 

We hope that our parametrisations, and the ones easily obtainable exactly with the 
same technique with forthcoming HERA data, will be useful for the QCD analysis of 
present and future experiments at the Tevatron and the LHC.

\section{Acknowledgments}
We want to thank Gregory Soyez for useful discussions on DGLAP equations.

\newpage


\end{document}